\documentclass[preprint]{aastex}

\slugcomment{ }

\shorttitle{12\micron\ Flare Observations}
\shortauthors{Jennings et al.}

\begin{document}
\bibliographystyle{apj}

%% LaTeX will automatically break titles if they run longer than
%% one line. However, you may use \\ to force a line break if
%% you desire.

\title{ Solar Magnetic Field Studies Using the 12-Micron Emission Lines. IV.
 Observations of a Delta-Region Solar Flare}

%% Use \author, \affil, and the \and command to format
%% author and affiliation information.
%% Note that \email has replaced the old \authoremail command
%% from AASTeX v4.0. You can use \email to mark an email address
%% anywhere in the paper, not just in the front matter.
%% As in the title, you can use \\ to force line breaks.

\author{Donald E. Jennings \altaffilmark{1} and Drake Deming}
\affil{Planetary Systems Branch, Goddard Space Flight Center, \\
    Greenbelt, MD 20771}

\and 

\author{George McCabe\altaffilmark{1,2}}
\affil{Institute for Astrophysics and Computational Science, \\
The Catholic University of America, Washington, DC 20064}

\and

\author{Pedro Sada\altaffilmark{1}}
\affil{Departmento de Fisica y Mathematicas,\\
 Universidad de Monterrey, San Pedro Garza Garcia,\\
 N.L. 66259, Mexico}

\and

\author{Thomas Moran\altaffilmark{2}}
\affil{Center for Solar Physics and Space Weather,\\
 The Catholic University of America, Washington, DC 20064}

%% Notice that each of these authors has alternate affiliations, which
%% are identified by the \altaffilmark after each name.  Specify alternate
%% affiliation information with \altaffiltext, with one command per each
%% affiliation.

\altaffiltext{1}{Visiting Astronomer, National Solar Observatory, Kitt Peak.
NSO is operated by AURA, Inc.\ under contract to the National Science
Foundation}
\altaffiltext{2}{and Goddard Space Flight Center, Greenbelt MD 20771}

%% Mark off your abstract in the ``abstract'' environment. In the manuscript
%% style, abstract will output a Received/Accepted line after the
%% title and affiliation information. No date will appear since the author
%% does not have this information. The dates will be filled in by the
%% editorial office after submission.

\begin{abstract}
We have recently developed the capability to make solar vector (Stokes
IQUV) magnetograms using the infrared line of MgI at 12.32\ \micron.
On 24 April 2001, we obtained a vector magnetic map of solar active
region NOAA 9433, fortuitously just prior to the occurrence of an M2
flare.  Examination of a sequence of SOHO/MDI magnetograms, and
comparison with ground-based H-alpha images, shows that the flare was
produced by the cancellation of newly emergent magnetic flux outside
of the main sunspot.  The very high Zeeman-sensitivity of the 12\
\micron\ data allowed us to measure field strengths on a spatial scale
which was not directly resolvable.  At the flare trigger site,
opposite polarity fields of $2700$ and $1000$ Gauss occurred within a
single $2$ arc-sec resolution element, as revealed by two resolved
Zeeman splittings in a single spectrum.  Our results imply an
extremely high horizontal field strength gradient ($5$ G/km) prior to
the flare, significantly greater than seen in previous studies.  We
also find that the magnetic energy of the cancelling fields was more
than sufficient to account for the flare's X-ray luminosity.
\end{abstract}

%% Keywords should appear after the \end{abstract} command. The uncommented
%% example has been keyed in ApJ style. See the instructions to authors
%% for the journal to which you are submitting your paper to determine
%% what keyword punctuation is appropriate.


%% From the front matter, we move on to the body of the paper.
%% In the first two sections, notice the use of the natbib \citep
%% and \citet commands to identify citations.  The citations are
%% tied to the reference list via symbolic KEYs. The KEY corresponds
%% to the KEY in the \bibitem in the reference list below. We have
%% chosen the first three characters of the first author's name plus
%% the last two numeral of the year of publication as our KEY for
%% each reference.

\section{Introduction}

Solar flares are believed to be powered by the release of magnetic
energy, but the details of the process are not understood.
Flare-related changes in magnetic configurations of active regions
have long been sought, without clear success.  However, recent
observations using ground-based \citep{cam99,har01} and space-borne
magnetographs \citep{kos99, kos01} are beginning to show clear
signatures of magnetic energy release, temporally coincident with
large flares.  It has long been known that small to moderate flares
can occur where newly-emergent flux becomes abutted against its
opposite polarity \citep{mar84} and is cancelled, presumably by
reconnection.  Flux-cancellation represents one obvious type of
flare-related magnetic change, but it is also believed that flux
cancellation can trigger the release of magnetic energy stored over
larger volumes \citep{pre84}. The derivation of the magnetic energy
from observations requires accurate measurement of the magnetic field
strength, and conventional magnetographs can give large errors in
field strength.  Because Zeeman splitting for visible-region lines is
less than the intrinsic line width, it can be difficult to
discriminate between a strong field having a small filling factor, and
a much weaker field with unit filling factor, and these two cases can
represent vastly different magnetic energies.

In recent years, infrared (IR) lines have been increasingly used
\citep{meu98}, because they can exhibit resolved Zeeman splitting,
giving the field strength independent of filling factor for fields
stronger than some limit (typically $\sim 400$ Gauss at 12\
\micron).  We have been developing instrumentation to make vector
field maps (Stokes $IQUV$) of solar active regions using the far-IR
line of MgI at 12.32\ \micron.  One of our vector field maps was
fortuitously made just prior to the occurrence of an M2 flare
initiated by flux cancellation.  Our data have insufficient temporal
resolution to demonstrate a decrease in magnetic energy at the precise
time of the flare.  Nevertheless, we exploit the great
Zeeman-sensitivity of these data to demonstrate that the flare began
where the horizontal gradient in field strength was exceptionally
large ($\gtrsim 5$ G/km), and that the magnetic energy in the
immediate vicinity of the flux-cancellation site was more than ample
compared to the flare's X-ray luminosity.

\section{Observations}

We observed the largest sunspot in NOAA region 9433 on 23-25 April
2001, at the McMath-Pierce telescope of the National Solar Observatory
on Kitt Peak.  We used our cryogenic grating spectrometer, `Celeste'
mounted on top of the main spectrograph tank to take advantage of its
ability to track the image rotation. Celeste achieves nearly
diffraction-limited spectral and spatial resolution, and
background-limited sensitivity.  It uses a large echelle grating ($18
\times 33$ cm$^{2}$) to achieve a resolution of 0.04 cm$^{-1}$ on the
12.32\ \micron\ line, corresponding to a magnetic resolution of $\sim
400$ Gauss. (We define magnetic resolution as the minimum field
strength difference which becomes visible as distinct Zeeman
splittings in a single spectrum. See \citet{dem91} for a discussion.)
As shown in Figure 1, the beam from the telescope is collimated,
passed through the polarimeter optics, and is focused into Celeste. In
the spectrometer, the beam passes through a filter wheel, and forms an
image at a $2$ arc-sec $\times\ 2.3$ arc-minute slit. The beam is then
collimated by internal Cassegrain optics, diffracted at the grating,
and refocused at the detector array.  The array is a $128
\times 128$ blocked-impurity-band device, with 75\ \micron\ pixels,
giving 1 arc-sec per pixel scale, double-sampling the $\sim 2$ arc-sec
telescope diffraction image. The spectrometer is housed in a
liquid-Helium dewar.

We record full vector magnetograms.  The mapping process consists of
stepping the solar image using a limb-guider, and cycling through the
Stokes parameters sequentially at each spatial position. The process
is performed at a steady cadence under computer control. In the
polarimeter optics the beam transits a 1/4-wave plate, a 1/2-wave
plate, and a chopping linear polarizer. By selecting combinations of
rotation angles for the two waveplates, followed by the chopping
polarizer, we measure $I$, $Q$, $U$, and $V$. We use a
double-differencing technique to subtract unpolarized light, and to
cancel any imbalance between the two polarizer positions.  Beginning
with the waveplate settings for $Q$, the two positions of the chopping
polarizer give spectra of $I_{u}+Q$ and $I_{u}-Q$.  The next waveplate
setting gives $I_{u}-Q$ and $I_{u}+Q$ (reverses the sign) for the two
polarizer positions. The sum of these yields $I_{u}$ and the
differences yield $Q$.  This cycle is repeated for $U$ and $V$.
Stokes $I$ is calculated from $I^{2}=Q^{2}+U^{2}+V^{2}$.  Thus all
four Stokes parameters, and also the unpolarized spectrum $I_{u}$ are
derived from the measurements.  This method of observing the
12.32\ \micron\ line Stokes parameters is an adaptation and extension of
the method pioneered by \citet{hew93}.

Figure 2 shows a SOHO Michaelson Doppler Imager (MDI) magnetogram and
continuum image of the observed spot, as well as an H-alpha image from
the Big Bear Solar Observatory (BBSO).  The immediate vicinity of the
observed (positive polarity) spot consisted of smaller pores of
negative polarity.  We inspected a sequence of MDI magnetograms from
$10$ hours before to several hours after our observations. This showed
that newly-emergent positive polarity flux was diverging from the
point marked with `+' on Figure 2.  Positive flux was also flowing
radially outward from the main spot, so flux emergent from the `+'
point was forced into close coincidence with the small negative
polarity spots in the two regions outlined with boxes on the upper
right panel of Figure 2. Opposite polarity umbrae existing within a
single sunspot are conventionally termed a `delta configuration.'
Although the opposite polarity regions observed here are not mature
umbrae, they do have field strengths attaining umbral values ($\gtrsim
2000$ Gauss).  We therefore refer to these two locations where
opposite polarities are in contact as `delta regions'.

During our 24 April IQUV mapping sequence an M2-class X-ray flare
occurred in the region. Relevant timing data for the flare and our
mapping sequence are given in Table 1.  H-alpha sequences recorded at
BBSO, and the H-alpha video recorded on the Razdow telescope at NSO
Kitt Peak, show a general brightening in the delta regions surrounding
the spot (see Figure 2, lower right panel), which increased gradually
in the hours before the flare.  The flare began with a sudden
additional H-alpha brightening in delta region 2, which spread rapidly
to region 1.  Clouds interfered toward the end of our mapping
sequence, but fortunately this happened after the flare onset, and
after the most significant portions of the 12\ \micron\ magnetogram
were complete.

Figure 3 shows an example of transverse ($(Q^{2}+U^{2})^{1/2}$, upper
panel) and longitudinal field (Stokes-V) images made during our
mapping sequence.  These are detector-plane images made at the slit
position marked on Figure 2.  One dimension is spatial, and cuts
through the main spot, extending into delta region number 1. The other
dimension is frequency in wavenumbers, and shows the familiar Zeeman
patterns for circular and linear polarization as image brightness
variations.  Because of the large splitting in this line, the
magnitude of the linear polarization is comparable to the circular
polarization.

%% In this section, we use  the \subsection command to set off
%% a subsection.  \footnote is used to insert a footnote to the text.

%% Observe the use of the LaTeX \label
%% command after the \subsection to give a symbolic KEY to the
%% subsection for cross-referencing in a \ref command.
%% You can use LaTeX's \ref and \label commands to keep track of
%% cross-references to sections, equations, tables, and figures.
%% That way, if you change the order of any elements, LaTeX will
%% automatically renumber them.

%% This section also includes several of the displayed math environments
%% mentioned in the Author Guide.

\section{Results}

The data shown here are among the first 12\ \micron \ polarization
maps ever observed for a solar active region.  Given the large Zeeman
sensitivity, and greater height of formation for this line, new
features are to be expected.

Prior to showing the 12\ \micron \ magnetic maps, we point out some
interesting aspects of the sample Stokes profiles illustrated on
Figure 3. Between $\sim 30$ and $\sim 70$ arc-sec, the slit crosses
the penumbra of the main spot. The Stokes-V profiles show large
apparent Doppler shifts, which reverse as the slit passes the center
of the spot. Since this sunspot was close to disk center (N18E06), the
direction of these flows is primarily upward and downward on opposite
sides of the spot.  This flow pattern is not uniquely related to the
flare, because it was also observed on April 23 and 25.  The apparent
magnitude of the flow is supersonic, $\sim 20$ km/sec. Although we
have observed several sunspots, this is the first instance of such
large apparent Doppler shifts.  Note that the transverse field does
not exhibit these shifts.  Although our method does not measure linear
polarization strictly simultaneously with circular, the same
properties (Doppler shifts in V, and not in Q and U) are seen over
extended regions of this sunspot on all days.  We tentatively
interpret this effect as resulting from the `fluted' nature of the
sunspot penumbra \citep{title93}, wherein field lines of drastically
different inclination can co-exist at the same general location.
Evidently we are seeing high-velocity flows along the field lines
which are nearly vertical, while simultaneously seeing linear
polarization from field lines which are nearly horizontal.  In
comparison to most lines used for magnetograms, the 12\ \micron\ line
is formed significantly higher (near $\sim 400$ km,
\citet{cha91,car92}) which may exaggerate effects due to sunspot
activity and fluting.

Near $\sim 85$ arc-sec the slit crosses delta region 1 (compare Figure
2). The field in this region is nearly all in Stokes-V; the transverse
component is negligible. Again, since this active region was close to disk
center, the fields in the delta region are nearly vertical,
i.e. directed radially outward from sun center. The field strength, as
seen by the splitting in Stokes-V, is at least as large as the strongest
penumbral field in the main spot (the 12\ \micron\ line disappears in
sunspot umbrae, so we cannot compare directly with the main spot umbral field).
In delta region 2 (not illustrated) some linear polarization signal
does appear, but the fields are still primarily vertical, and their
field strengths are even greater.

Since the delta region fields are represented primarily in Stokes-V,
we use Stokes-V to illustrate our magnetic map of this region. In each
spectrum, we determined the zero-crossing wavelength of the V-profiles
(using spline interpolation), and we `folded' the profiles about their
zero-crossing point.  Since our Stokes-V profiles are not
systematically asymmetric, the folding increases the signal-to-noise
level without losing physical information. Because the Zeeman
sensitivity of this line is so large, we can show magnetograms {\em at
a given field strength}.  For example, only weak fields contribute
near the center frequency of the line, because the large Zeeman
splitting from strong fields displaces their Stokes-V components away
from line center.  So the third dimension of our magnetic map is field
strength.  Figure 4 shows our magnetogram `sliced' at 350, 700, 1400,
2100, 2800, and 3900 Gauss. Near 80 arc-sec on the horizontal scale
the data were affected by clouds, but we verified that the important
delta region 2 near 70 arc-sec was unaffected.

\section{Discussion}

The 350 Gauss panel of Figure 4 shows a noticeably more `diffuse'
appearance than the panels at higher field strengths.  This shows that
weak fields are present in active regions. They are broadly
distributed, in contrast to strong fields - which concentrate at
specific locations.  A large fraction of the weak field component is
due to the sunspot `canopy', which can be seen as the extended region
of positive polarity most obvious in Stokes-V on the side of the
sunspot toward disk center (compare Figure 2). We have previously
described a prominent sunspot canopy observed in the 12.32\
\micron\ line \citep{jen01}.

The sunspot penumbral field shows a ring-like morphology, which of
course reflects the fact that the single field strength in each panel
is found at only a particular radius from the spot center.  With
increasing field strength, the width and radius of the ring decrease.
At the largest field strength (3900 Gauss), the polarity of the
sunspot field seems to reverse.  We have observed this effect in
several sunspots, and it is certainly not an observational artifact.
However, it does not represent an actual polarity reversal.  Recall
that this emission line is flanked by a broad, shallow, absorption
trough \citep{chasch91}.  At large distances from line center, the
sunspot Stokes-V profile is dominated by the relatively low-amplitude
V signal from the absorption trough, which accounts for the apparent
polarity reversal.  In principle, this effect conveys information on
the height gradient of field strength, since the absorption is formed
deeper than the emission.  However, a quantitative interpretation is
beyond the scope of this paper.

The MDI whole-disk magnetogram at 17:36 UT registers peak field
strengths of $\sim 250$ and $\sim 500$ Gauss for the positive polarity
fields in delta regions 1 and 2 respectively.  In contrast, the 12\
\micron\ data show that these regions are not even prominent in
positive polarity until the field strength exceeds $700$ Gauss, with
significant intensity being recorded even at $2800$ Gauss.  Evidently
the fields are not resolved at the MDI whole disk resolution of $2$
arc-sec.  Although our spatial resolution was also $2$ arc-sec, the
very large Zeeman splitting in the 12\ \micron\ line permits us to
measure true field strengths without spatially resolving the
fields. In both delta regions 1 and 2 the average strength of the
positive polarity fields is $\sim 2000$ Gauss, but in delta region 2
(where the flare was triggered), the distribution of field strengths
extends to greater values than in region 1. This is noticeable in the
3900 gauss panel, where a trace of delta region 2 can still be
discerned.  Moreover, these fields are in quite close proximity to the
small negative polarity pores, producing exceptionally large field strength
gradients.  Figure 5 shows the Stokes-V profile at the point in delta
region 2 where opposite polarities are in contact (marked by the
circle on the 2100 Gauss panel).  Within the $2$ arc-sec spatial
resolution element, two distinct Zeeman splittings are seen: a 2700
Gauss splitting at positive polarity, and 1000 Gauss at negative
polarity. These splittings are marked on Figure 5.  We contemplated
alternative interpretations of this complex Stokes-V profile, which
might lessen the necessity for strong opposite-polarity fields in
close spatial coincidence. However, alternate interpretations are
difficult to reconcile with the Stokes-V profiles in immediately
adjacent regions, which support the picture of strong fields in both
polarities.

This large field strength difference in such close proximity requires
a horizontal field gradient of $\sim 5$ G/km.  To our knowledge, this
is the largest horizontal gradient ever observed on the Sun, and suggests
reconnection at this site, at or near the temperature-minimum height.
In their study of upper-photospheric reconnection, \citet{litmar99}
remark that `the local field cannot be...measured directly.' The high
magnetic resolution afforded by the 12\ \micron\ line indeed makes
such a measurement possible, and the height of line formation is well
matched to the temperature-minimum region where the electrical
resistivity is maximum.  We therefore point out the significant
diagnostic potential of the 12\ \micron \ line in studies of
photospheric reconnection.

Other investigators find high field strength gradients at flaring
sites, but less than we infer here. \citet{kos01} found a gradient of
$1.3$ G/km for the X5.7 `Bastille Day' flare; our results for a
significantly weaker flare suggest that conventional magnetographs
(even space-borne ones such as SOHO/MDI) underestimate the field
strength gradients.  If so, they probably also underestimate the
magnitude of magnetic energy release inferred from observed changes in the
fields.

Our results have significant implications for the magnetic energy
available locally (i.e., in the flux-cancellation region of the mid-
to upper photosphere) to power this flare.  The total magnetic energy
is proportional to the volume integral of $\frac{B^{2}}{8\pi}$
\citep{chan}.  We estimate this quantity directly from
the Stokes-V profiles:

\begin{equation}
E_{tot} = \frac{\delta H}{8\pi} \epsilon \sum B^{2} f_{B} dA
\end{equation}

where the volume integral is approximated as a discrete sum over field
strengths within a spatial resolution area $dA$, times a layer
thickness $\delta H$. $\epsilon$ is the fraction of the atmosphere
which is magnetic, i.e. a filling factor.  We constrain $\epsilon >
0.15$ by comparing the MDI and 12\ \micron\ field strengths. The
factor $f_{B}$ expresses the fraction of the magnetic atmosphere which
exhibits field strength $B$.  The essential advantage of a
strongly-split infrared line is that we can infer $f_{B}$ as
proportional to the amplitude of the Stokes-V profile at each
splitting (i.e., field strength) value.  (In so doing we assume that
the line formation mechanism for the 12.32\ \micron\ transition is
independent of field strength.)  Moreover, because the 12\ \micron\
line is formed in the temperature minimum region, $\sim400$ km above
the photosphere \citep{cha91, car92}, we know that $\delta H > 400$
km.  On this basis we find that the total `local' magnetic energy of
region 1 was $> 3 \times 10^{22}$ Joules, and of region 2, $> 4.5
\times 10^{22}$ Joules.  Using the single-value MDI field strengths
with unit filling factor gives $3.5 \times 10^{20}$ and $6 \times
10^{21}$ Joules for regions 1 and 2 respectively. By comparison, the
X-ray luminosity of the flare was $3 \times 10^{21}$ Joules.  As is
well known, only a fraction of the total magnetic energy is available
to a flare \citep{low90}, and the flare's total luminosity will of
course be greater than the energy measured in the GOES X-ray bands
(Table 1).  Hence the MDI energies are uncomfortably small, but the
12\ \micron\ field strengths indicate an ample local resevoir of
magnetic energy.  A conventional view is that cancelling magnetic flux
simply triggers the release of magnetic energy stored in much larger
volumes.  While this may actually happen for many of the largest
flares, our results imply that sufficient energy was available to
power this M2 flare `locally', without invoking the release of
magnetic energy from much larger volumes.

%% If you wish to include an acknowledgments section in your paper,
%% separate it off from the body of the text using the \acknowledgments
%% command.

%% Included in this acknowledgments section are examples of the
%% AASTeX hypertext markup commands. Use \url without the optional [HREF]
%% argument when you want to print the url directly in the text. Otherwise,
%% use either \url or \anchor, with the HREF as the first argument and the
%% text to be printed in the second.

\acknowledgments

We thank the referee, Dr. William Livingston, for comments which
improved this paper.  This work was facilitated by the open data
access policy of the SOHO/MDI instrument, and an H-alpha image from
Big Bear Solar Observatory. BBSO is operated by the New Jersey
Institute of Technology, through funding by the NSF and NASA. We are
grateful to Jack Harvey for communicating the content of his AGU talk,
and for helping with access to some GONG+ magnetograms.  Celeste was
manufactured by IR Systems, and its electronics were manufactured by
Wallace Instruments.  The detector array was produced by Boeing
Research and Technology Center.

\clearpage

%% Use the figure environment and \plotone or \plottwo to include 
%% figures and captions in your electronic submission.

\begin{figure}
\epsscale{0.5}
\plotone{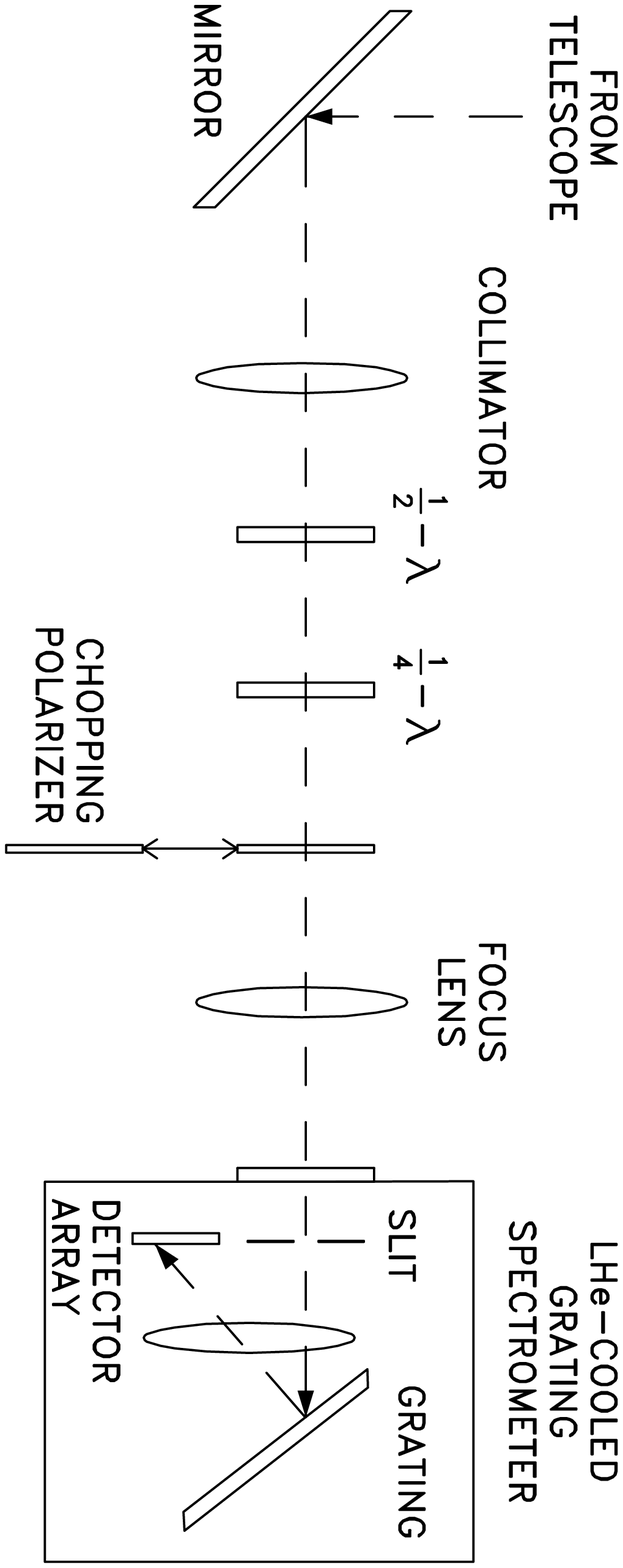}
\vspace{0.5 in}
\caption{Optical layout used to measure Stokes $IQUV$ profiles at 12.32\ \micron.
 \label{fig1}}
\end{figure}

\clearpage

\begin{figure}
\epsscale{1.0}
\plotone{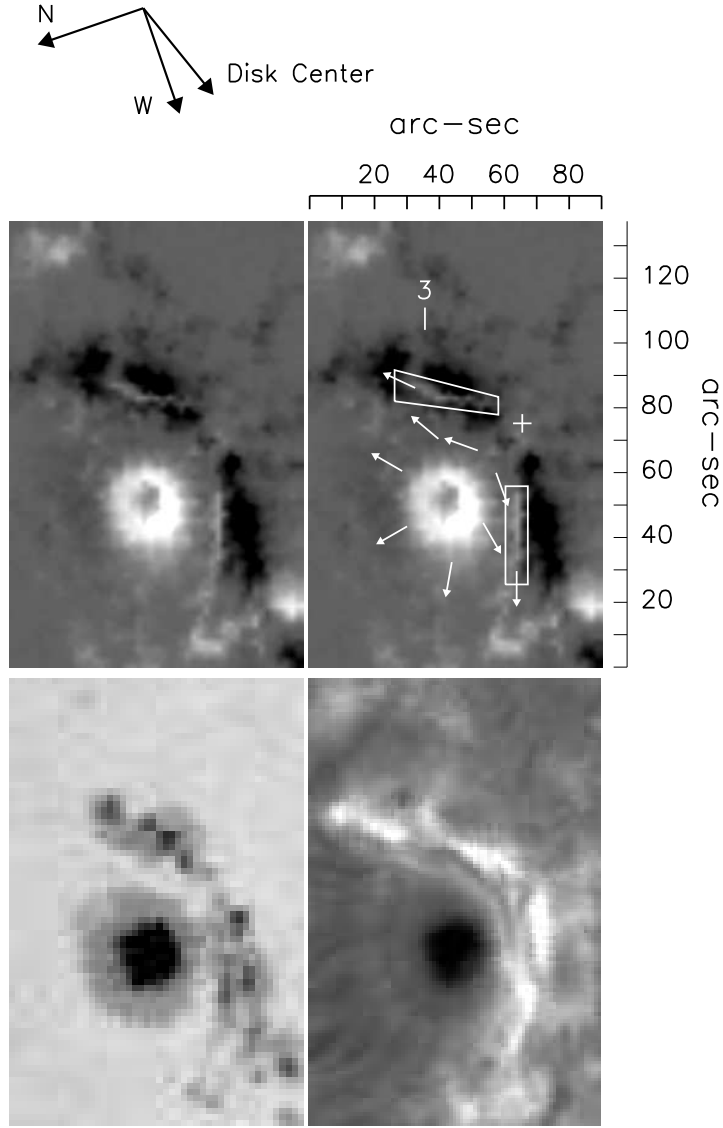}
\caption{MDI magnetogram (upper left, 17:36 UT), white light (lower left,
17:18 UT, and BBSO H-alpha (lower right, 17:58 UT) images of the main spot
in NOAA 9433 taken just prior to our 12\ \micron\ IQUV data.  The
upper right panel repeats the MDI magnetogram, and superposes arrows
which indicate the direction of flux transport (length of the arrows
has no significance). Moving positive flux (brighter areas) diverges
from the spot marked with `+'. The two boxed areas indicate `delta
regions' having opposite polarity in close proximity; the upper we
designate as region 1, and the lower right one as region 2.  The area
and orientation of these images were adjusted to match our 12\
\micron\ data (Figure 4). The line marked `3' indicates the slit
position corresponding to Figure 3; the 12\ \micron \ slit was
oriented vertically, and stepped to the right. \label{fig2}}
\end{figure}

\clearpage

\begin{figure}
\plotone{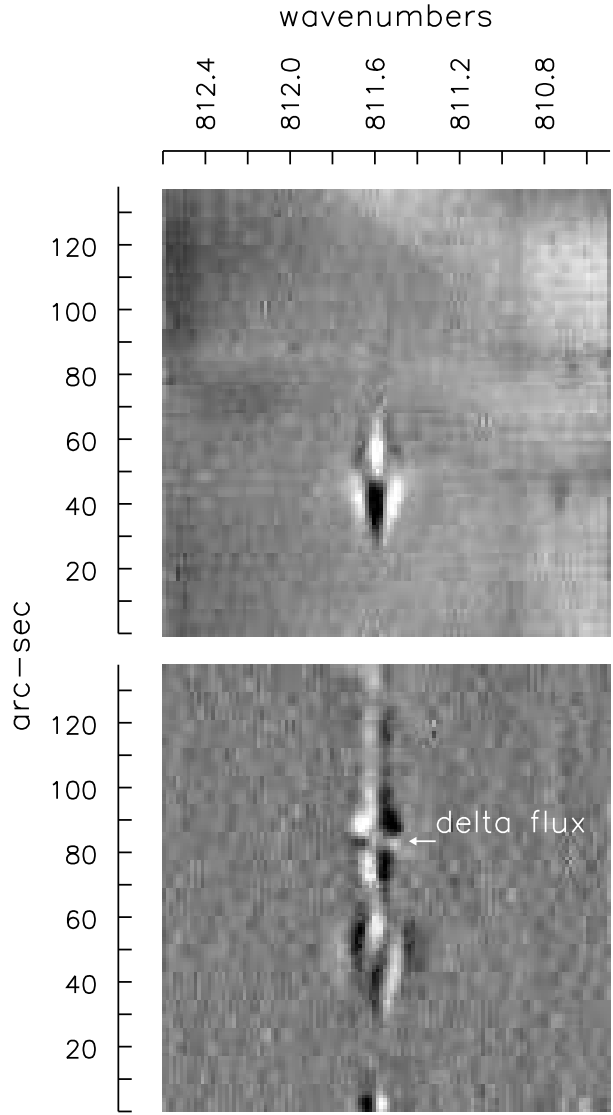}
\caption{Sample Stokes images for a slit position cutting through the penumbra
 of the main spot, and into the delta configuration.  The top image is
 the total linear polarization, i.e. $(Q^{2}+U^{2})^{1/2}$ (we measure
 Q and U separately), and the bottom image is Stokes-V.  The reversal
 of polarity in the delta region number 1 (see Figure 2) is marked by
 an arrow.  A mark on Figure 2 identifies the location of the slit for
 these sample profiles. \label{fig3}}
\end{figure}

\clearpage

\begin{figure}
\epsscale{0.8}
\vspace{0.5 in}
\plotone{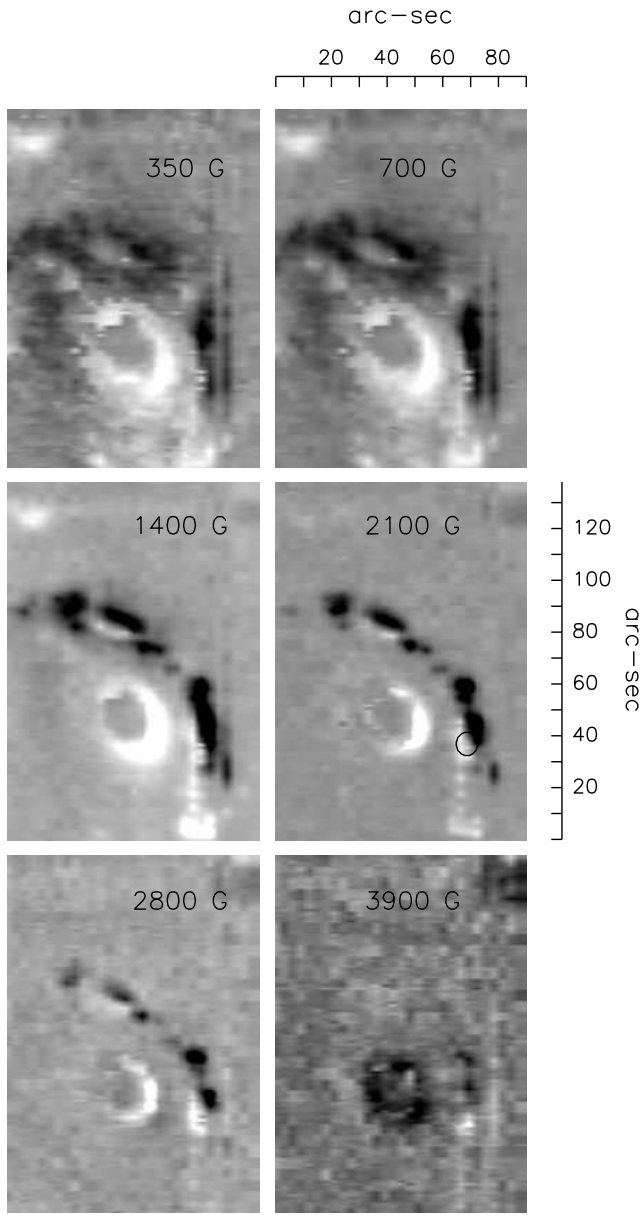}
\vspace{1.5 in}
\caption{12\ \micron\ magnetograms made by `slicing' the Stokes-V image
cube at increasing values of the field strength.  Compare to the MDI
magnetogram (Fig 2).  Brightness indicates the polarity and intensity
of the line emission at the splitting value corresponding to the given
field strength (positive polarity is bright, negative is dark).  The
intensity scale of each image is normalized separately, for clarity of
display.  These magnetograms were obtained by scanning the Celeste
slit (vertical on this representation) from left to right. The
Stokes-V profile in the region marked by a circle on the 2100 Gauss panel is
plotted on Figure 5. \label{fig4}}
\end{figure}

\begin{figure}
\epsscale{1.0}
\plotone{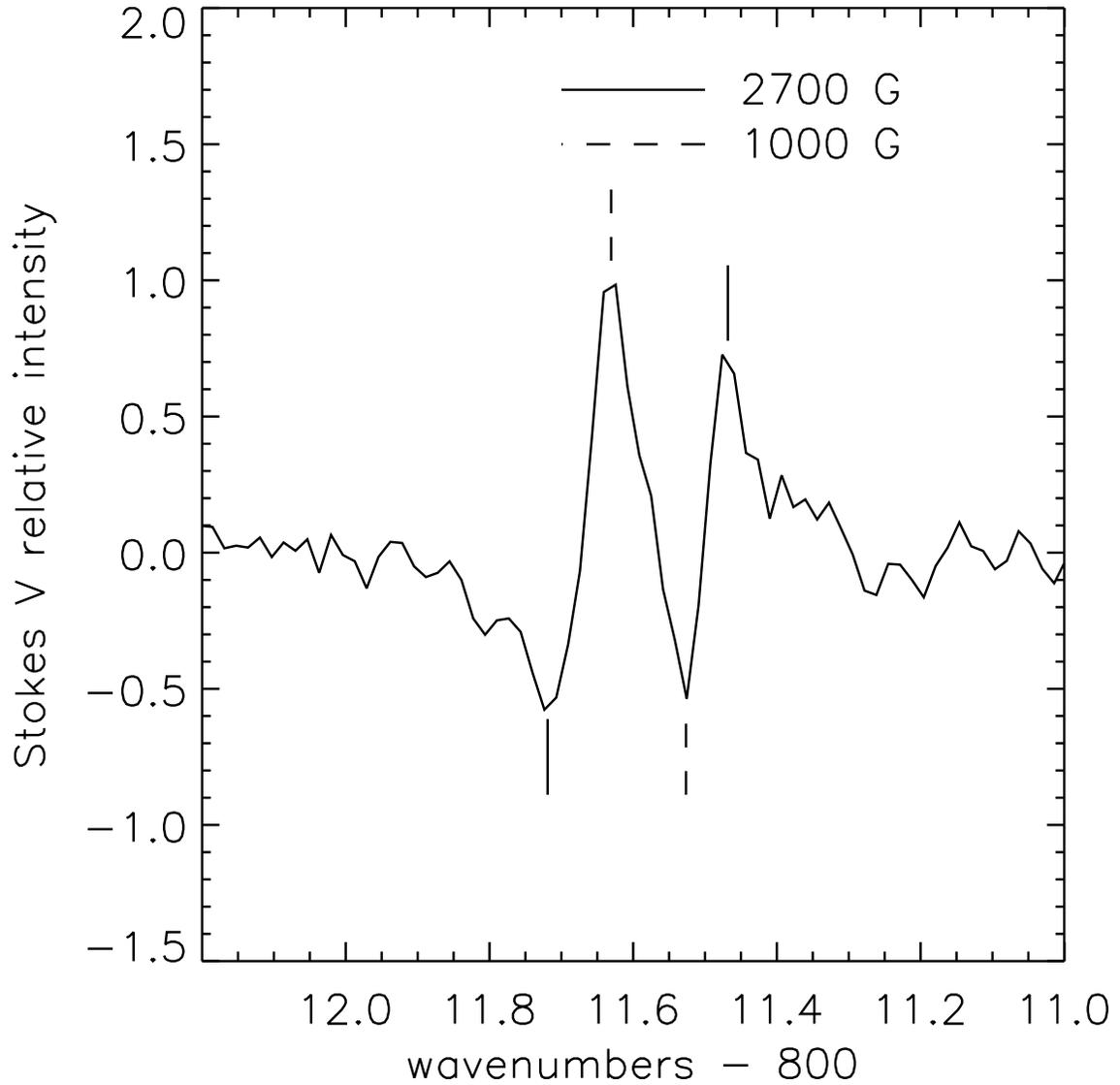}
\caption{ Stokes-V profile for the specific location in delta region 2
 (marked with a circle on Figure 4) where strong-field opposite
 polarities are in close coincidence.  Two separate Zeeman splittings
 of 2700 and 1000 Gauss, with opposite polarities, are
 marked. Line center is at 811.58 cm$^{-1}$. \label{fig5}}
\end{figure}

\clearpage

%% If you are not including electonic art with your submission, you may
%% mark up your captions using the \figcaption command. See the 
%% User Guide for details.
%%
%% No more than seven \figcaption commands are allowed per page, 
%% so if you have more than seven captions, insert a \clearpage 
%% after every seventh one. 

%% Tables should be submitted one per page, so put a \clearpage before
%% each one.

%% Two options are available to the author for producing tables:  the
%% deluxetable environment provided by the AASTeX package or the LaTeX
%% table environment.  Use of deluxetable is preferred.
%%

%% Three table samples follow, two marked up in the deluxetable environment,
%% one marked up as a LaTeX table.

%% In this first example, note that the \tabletypesize{}
%% command has been used to reduce the font size of the table.
%% Note also that the \label command needs to be placed 
%% inside the \tablecaption.

\clearpage

\begin{deluxetable}{lr}
\tablecaption{Parameters for the Flare and 12\ \micron\ Map\label{tabpar}}
\tablewidth{0pt}
\tablehead{
\colhead{Parameter} &
\colhead{   }}
\startdata
Flare Begin & 18:04 UT \\ Flare Peak & 18:12 UT \\ Flare End & 18:17
UT \\ Flare X-ray Luminosity & $3 \times 10^{21}$ Joules \\ 12\ \micron\
Map Start & 16:15 UT \\ 12\ \micron\ Map End & 18:52 UT \\
\enddata
\end{deluxetable}

%% If you use the table environment, please indicate horizontal rules using
%% \tableline, not \hline.
%% Do not put multiple tabular environments within a single table.
%% The optional \label should appear inside the \caption command.

\clearpage


\begin{thebibliography}{}
\bibitem[Cameron and Sammis (1999)]{cam99} Cameron, R., and Sammis, I.
	 1999, \apjl, 525, L61
\bibitem[Carlsson et al. (1992)]{car92} Carlsson, M., Rutten, M. J.,
	  and Shchukina, N. G. 1992, \aap, 253, 567
\bibitem[Chandrasekhar(1961)]{chan}  Chandrasekhar, S.  1961, Hydrodynamic
  and Hydromagnetic Stability, Dover Publications.
\bibitem[Chang et al.(1991)]{cha91} Chang, E. S., Avrett, E. H.,
	 Mauas, P. J., Noyes, R. W., and Loeser, R.  1991, \apjl, 379, L79
\bibitem[Chang and Schoenfeld (1991)]{chasch91} Chang, E. S.,
	 and Schoenfeld, W. G.  1991, \apj, 383, 450
\bibitem[Deming et al.(1991)]{dem91} Deming, D., Hewagama, T., Jennings, D. E.,
	 and Wiedemann, G. 1991, in Solar Polarimetry, Proceedings of
	 the Eleventh National Solar Observatory / Sacramento Peak
	 Summer Workshop, L. J. November (ed), National Solar
	 Observatory, Sunspot NM, p.341.
\bibitem[Harvey (2001)]{har01} Harvey, J. W., 2001, American
	Geophysical Union, Spring Meeting 2001, abstract SH22A-01
\bibitem[Hewagama et al. (1993)]{hew93} Hewagama, T., Deming, D.,
	 Jennings, D. E., Osherovich, V., Wiedemann, G., Zipoy, D.,
	 Mickey, D. L., and Garcia, H. 1993, \apjs, 86, 313
\bibitem[Jennings et al.(2001)]{jen01} Jennings, D. E., Deming, D.,
	 Sada, P. V., McCabe, G. H., and Moran, T.  2001, in Advanced
	 Solar Polarimetry: Theory, Observation and Instrumentation,
	 M. Sigwarth (ed), ASP Conference Series 236, 273
\bibitem[Kosovichev and Zharkova (1999)]{kos99} Kosovichev, A. G.,
    and Zharkova, V. V.  1999, \solphys, 190, 459
\bibitem[Kosovichev and Zharkova (2001)]{kos01} Kosovichev, A. G.,
    and Zharkova, V. V.  2001, \apjl, 550, L105
\bibitem[Litvinenko and Martin (1999)]{litmar99} Litvinenko, Y. E.,
	and Martin, S. F. 1999, \solphys, 190, 45
\bibitem[Low and Lou (1990)]{low90} Low, B. C., and Lou, Y. Q. 1990,
    \apj, 352, 343
\bibitem[Martin et al.(1984)]{mar84} Martin, S. F., and 10 co-authors
	1984, Adv. Space Res., 4, 61
\bibitem[Meunier et al. (1998)]{meu98} Meunier, N., Solanki, S. K., 
    and Livingston, W. C.  1998, \aap, 331, 771
\bibitem[Priest (1984)]{pre84} Priest, E. R.  1984, Adv. Space Res., 4, 37
\bibitem[Title et al. (1993)]{title93} Title, A. M., Frank, Z. A.,
	 Shine, R. A., Tarbell, T. D., Topka, K. P., Scharmer, G.,
	and Schmidt, W. 1993, \apj, 403, 780
\end{thebibliography}
\end{document}